\documentclass[aps,nofootinbib,showpacs,twocolumn,longbibliography]{revtex4-1}
\usepackage{amstext,amsmath,amssymb,amsfonts,bbm}
\usepackage[latin1]{inputenc}
\usepackage{graphicx}
\usepackage{epstopdf}
\usepackage{amsthm}
\usepackage{tocvsec2}
\usepackage{enumerate}
\usepackage{subfigure}
\usepackage{color}

\usepackage{multirow} \usepackage{rotating}
\usepackage[colorlinks,citecolor=blue,linkcolor=blue,urlcolor=blue]{hyperref}

\def\beq{\begin{equation}}
\def\be{\begin{equation}}
\def\ee{\end{equation}}
\def\ba{\begin{eqnarray}}
\def\ea{\end{eqnarray}}

\begin{document}

\title{Coupling and thermal equilibrium in general-covariant systems}

\author{Goffredo Chirco}
\author{Hal~M.~Haggard}
\author{Carlo Rovelli}
\affiliation{Aix Marseille Universit\'e, CNRS, CPT, UMR 7332, 13288 Marseille, France.}
\affiliation{Universit\'e de Toulon, CNRS, CPT, UMR 7332, 83957 La Garde, France.}

\date{\today}
\begin{abstract}
\noindent A fully general-covariant  formulation of statistical mechanics is still lacking. We take a step toward this theory by studying the meaning of statistical equilibrium for coupled, parametrized systems. We discuss how to couple  parametrized systems. We express the thermalization hypothesis in a general-covariant context.  This takes the form of vanishing of information flux.  An interesting relation emerges between thermal equilibrium and gauge.
\end{abstract}
\pacs{04.20.Cv,04.20.Fy,05.20.Gg,05.70.-a}
\maketitle

\section{Introduction}

Gravity, thermodynamics and quantum mechanics form a knot where much of what we do not  yet understand about the world appears to hide.  Here we focus on the relation between the first two strands of the knot, gravity and thermodynamics. 

It has been repeatedly pointed out that while statistical mechanics on a given curved spacetime is fairly clear, a theory of the full statistical mechanics of the gravitational field is not yet available.  We understand the statistical fluctuations of the electromagnetic field (for instance with black-body theory), but not those of the gravitational field,  beyond the linear approximation. We think that this is one of the reasons for the puzzling aspects in several of the current speculations on the relation between thermodynamics and gravity \cite{ja95,japa,car,pad11,pad05}. Standard thermodynamics and statistical mechanics are based on notions (such as a preferred time) which have no equivalent in a general-covariant theory, where coordinate-time evolution is gauge and there is no preferred time flow. What is equilibrium in this context? What is thermalization?  What is equipartition of energy, if  gravitational energy is such a slippery concept as it is in general relativity? 

A line of thinking on these problems, and one that works towards constructing a coherent general-covariant statistical mechanics, is based on the idea of thermal time \cite{ro93, ro93_2,roco, robook, cobook}.  The idea is to reinterpret the relation between  time flow (generated by the Hamiltonian $H$) and Gibbs states $\rho\propto e^{-\beta H}$, by viewing the first as being determined by the second rather than the second determined by the first.  The time flow with respect to which a covariant state is in equilibrium can be read out from the state itself, and the germ of temporality is traced to quantum non-commutativity \cite{cobook}. 

The main problem that this approach leaves open is characterizing equilibrium, distinguishing thermal states from general statistical ones.\footnote{An attempt of doing so is in \cite{ro12}, where physical equilibrium states are understood as those whose thermal time is a flow in space-time. 
A generalization of the statistical derivation of the uniformity of temperature in the relativistic context was given in \cite{hal} for the case of equilibrium in a stationary space-time. A derivation \`{a} la Boltzmann of the distribution on the space of timeless noninteracting states was given in \cite{monte}.}  Here we address this problem by going back to basics and studying the notion of equilibrium for two coupled general-covariant systems.

Coupling generally-covariant systems is subtle, as we show below.  We study how two generally covariant system can couple in Section \ref{cl}, after a brief review of the Hamiltonian description of general covariant systems \cite{dirac}; this opens a wealth of interesting questions.  Once these have been clarified, we study equilibrium in the context of this coupling in Section \ref{see}, using the idea of thermal time. 

We obtain two surprising results. The first is that  we recover directly the characterization of equilibrium as vanishing information flow introduced in \cite{hal}.  The second is a tantalizing relation between equilibrium and gauge, which, in our opinion, provides strong support for the interpretation of gauge in terms of the unpredictable but 
measurable partial observables discussed in \cite{cr1} and \cite{gauge}.  

We summarize our results and the ensuing general picture of coupling and equilibrium for covariant systems in  Section \ref{disc}.

\section{General Covariant  Coupling} \label{cl}

\subsection{Dirac's generalized Hamiltonian dynamics} \label{gcl}

We start with a condensed review of Dirac's generalized Hamiltonian dynamics \cite{dirac}, with an example followed by the general case translated into modern language. 
A general-covariant system is defined by a Lagrangian that leads to a vanishing canonical Hamiltonian. Its  equations of motion are  invariant under reparametrization of the evolution parameter. The Legendre transform of the Lagrangian of these systems defines a phase space with constraints, and the dynamics is coded in the constraints.

One view on the goal of this formalism is that it expresses dynamics in a relational language \cite{ro12}. The physical correlations among dynamical variables are defined without specifying one of these as the independent ``time" variable. 

As an illustrative example consider a simple harmonic oscillator and its covariant Lagrangian 1-form
\be
L dt=  \frac{1}{2} \left((dq/dt)^2-q^2 \right)dt=\frac{1}{2} \left( \dot{q}^2/\dot{t}^2-q^2 \right)d \lambda ,
\ee
with position $q$, time $t$ and where dot denotes derivatives with respect to an arbitrary parameter $ \lambda $. Legendre transforming yields an extended phase space $\Gamma_{ex}$ with coordinates $(q^a, p_a) = (q,t,p,-E)$ where $a = 1, 2$, and the conjugate coordinates are interpreted as momentum $p$ and energy $E$. The symplectic form on $\Gamma_{ex}$ is  $\omega = dp_a \wedge dq^a $. The Hamiltonian vanishes identically and hence leads to the Hamiltonian constraint 
\be
C = E- \frac{1}{2}(q^2 +p^2) = E-H(p,q). 
\ee

The constraint surface can be coordinatized by $(q, t, p)$ and we find the restricted (pre-)symplectic form $\omega'=dp\wedge dq-dH\wedge dt$ on this surface. The flows corresponding to vectors in the kernel of the form $\omega'$ are called the orbits and they capture the physical correlations between the observables. These can be coordinated by the values $(q_0,p_0)$ at $t=0$, or, better, by the amplitude and phase $(A,\phi)$ that label the orbits 
\be
A=\sqrt{q^2 +p^2},\ \ \  \phi=\arcsin(q/\sqrt{q^2+p^2})-t \label{nn}.
\ee

The equations of \eqref{nn} are the relations between partial observables predicted by the theory. They can be rewritten in the recognizable  form 
\be
q = A\,\sin(t + \phi),\ \ \  \ p = A\, \cos(t +\phi). 
\ee 
Notice that this is a formulation of the dynamics of an oscillator where $q$ and $t$ are treated on an equal footing. The symplectic form can be further reduced to live just on the space of the orbits and a direct computation gives $\sigma = A \,dA \wedge d\phi$. (That is $\{A,\phi\}=A^{-1}$). 

\begin{figure}[t]
\includegraphics[width=0.4 \textwidth]{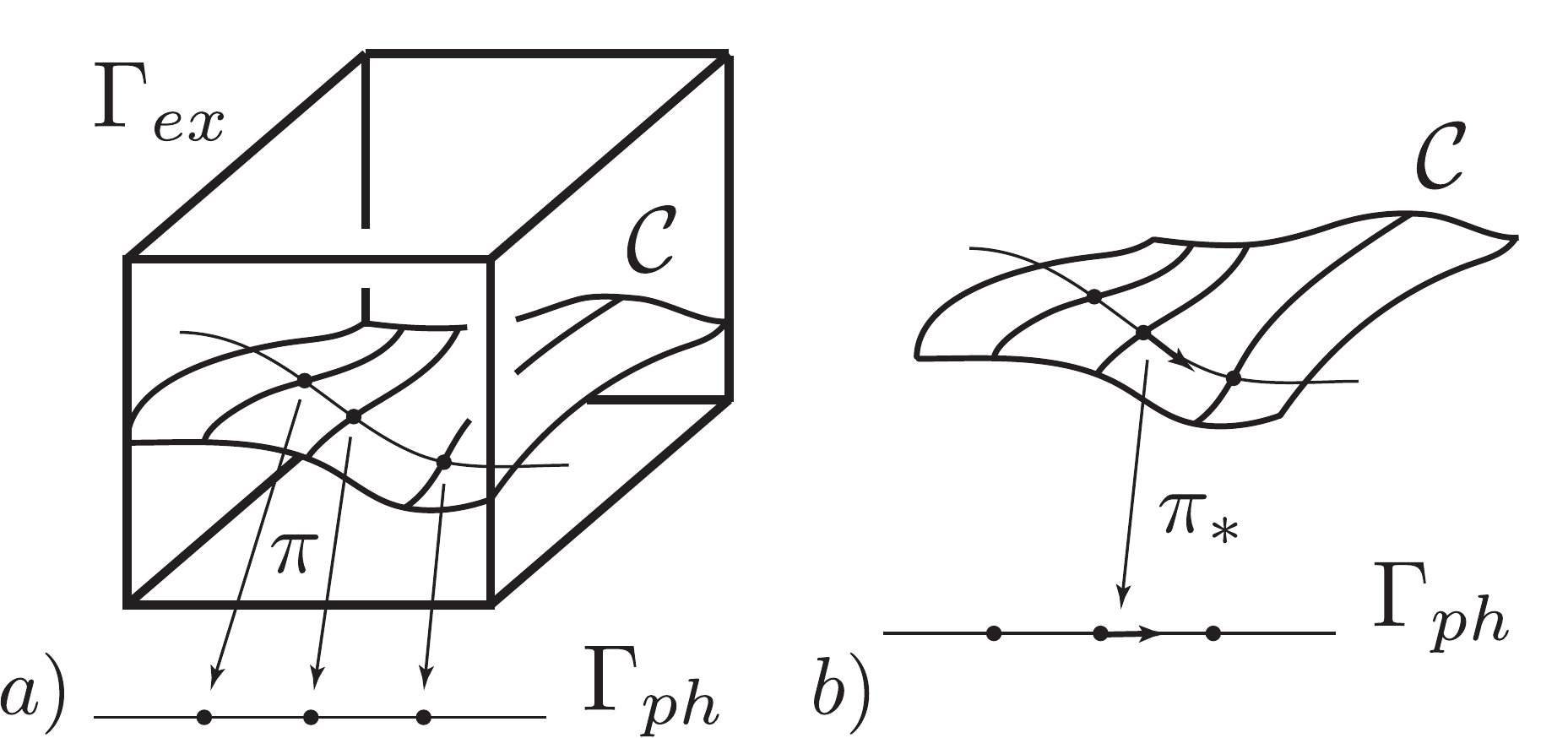}
\caption{Structure of a general covariant phase space. a) The extended phase space $\Gamma_{ex}$ and the constraint surface ${\cal C}$, which is foliated by orbits. The orbits project onto points of the physical phase space $\Gamma_{ph}$ under the projection $\pi$. b) The projection $\pi$ is also used to pull back forms to ${\cal C}$ and to push forward vector fields to the physical phase space. }\label{uno}
\end{figure}

We turn now to the general case, see e.g. \cite{robook}. Let $\Gamma_{ex}$ denote the (extended) phase space (cube of Fig.~\ref{uno}a) and $\mathcal{C}$ denote the subspace of $\Gamma_{ex}$ where the constraints vanish (surface of Fig.~\ref{uno}a). The symplectic space $\Gamma_{ex}$ has a canonical symplectic form $\omega$. The restriction of $\omega$ to $\mathcal{C}$, namely its pull back under the embedding $i$ of $\mathcal{C}$ into $\Gamma_{ex}$, is a pre-symplectic two-form $\omega'$. The space of its orbits (the integral surfaces of the $Y_i$ where $\omega'(Y_i ) = 0$ with  $ i=1, \dots, \dim{\ker \omega' \equiv D}$) is the physical phase space $\Gamma_{ph}$ (projected space of Fig.~\ref{uno}a), and carries the symplectic structure $\sigma$ defined by $\pi^{*}\sigma=\omega'$, where $\pi$ is the projection onto the orbit space.

Each point in $\Gamma_{ph}$ is a ``motion", namely a physically distinct solution of the equations of motion of the system, and establishes a system of relations (defined by its orbit) among the partial observables \cite{cr1}, which are  functions on $\Gamma_{ex}$ (restricted to $\mathcal{C}$). These relations are the predictions of the theory.

This formulation is  a generalization of conventional mechanics because it remains valid also for systems (like the relativistic ones), where $C$ does not have the form $E-H(q,p)$.

\subsection{Coupling} 

Two non-relativistic systems $S_1$ and $S_2$, with phase spaces $\Gamma_1$ and $\Gamma_2$ and Hamiltonians $H_1$ and $H_2$ can be given a {unified description} in terms of the coupled phase space $\Gamma=\Gamma_1\times\Gamma_2$, with Hamiltonian $H=H_1+H_2$. This kinematical coupling allows the possibility of a dynamical coupling between the systems, for instance by adding to $H$ a term, $V_{\text{int}}$, that does not factorize into the sum of two functions, one on $\Gamma_1$ and one on $\Gamma_2$. 

Can the same be done for general-covariant systems? Consider two {general covariant} systems $S_1$ and $S_2$, with (extended) phase spaces $\Gamma_{ex}^1$ and $\Gamma_{ex}^2$ whose dynamics are determined by the Hamiltonian constraints $C_1=0$ and $C_2=0$. {How do we couple them?}  

The obvious way is to consider the (extended) phase space  $\Gamma_{ex}=\Gamma_{ex}^1\times\Gamma_{ex}^2$, with symplectic form  $\omega=\omega_1 + \omega_2$ and the constraint system
\be
C_1=0, \ \ \ \ \ C_2=0.
\label{cs}
\ee  
This is simple and, we argue, correct, but subtle; to see this, consider an example. 

Let $\Gamma_{ex}^1$ admit canonical coordinates $(q_1,t_1,p_1,-E_1)$ and $C_1=E_1-H_1(p_1,q_1)$. This is the general-covariant form of a system with one degree of freedom and Hamiltonian $H(q,p)$. Say this is the covariant description of a pendulum.  Similarly for $(q_2,t_2,p_2,-E_2)$ and $C_2=E_2-H_2(p_2,q_2)$. The surface where the constraints are satisfied admits coordinates $(q_1,p_1,t_1,q_2,p_2,t_2)$. The constraint orbits are two-dimensional, and $(t_1,t_2)$ can be taken as parameters along each orbit. The physical phase space (the space of the orbits) can be coordinatized by the values of $(q_1,p_1,q_2,p_2)$ at $t_1=t_2=0$; this is fine.  But the dynamics is now curiously characterized by \emph{two}  times. What is the interpretation of this fact?  

A dynamical system is interpreted when we give a physical interpretation to its partial observables \cite{cr2}. The motion of the system is then described by the correlations between these partial observables. In our example, $(q_1,t_1)$ represents an event where the first pendulum is in a certain position $q_1$ of space at a certain value $t_1$ of time, measured with the clock $t_1$, and the same for $(q_2,t_2)$ and the other pendulum.  Once we fix a point in the physical phase space, that is a state of motion or a solution of the equations of motion, the theory gives us the position of the first pendulum at any $t_1$ and, independently, the position of the second at any $t_2$, for any arbitrary choice of $t_1$ \emph{and} $t_2$.  This is the simple physical interpretation of the double time parameter.   Correspondingly, each orbit in the constraint surface is two-dimensional. 

But, of course, we may not be interested in this. We may be interested uniquely in the position of the two oscillators at the same time  
\be
\label{syncro}
t=t_1=t_2. 
\ee 
For instance, we may assume that we are describing a setting in which our partial observables are only three: the two positions and a single clock. Then we only look at the position of both particles at a single clock time.  The theory provides this information, of course.  If we want to restrict ourselves to \emph{this} information, then we can pose equation \eqref{syncro} as a gauge fixing condition, which selects a one-dimensional trajectory within each 
two-dimensional orbit.  

The gauge fixing \eqref{syncro} splits the two-dimensional first-class constraint system \eqref{cs}  into two parts: the single constraint 
\be
C = C_1+C_2=0
\ee
that commutes with the gauge-fixing condition 
\be
\chi = t_2-t_1=0
\label{gf}
\ee
and the constraint 
\be
\Delta=C_1-C_2
\ee
that becomes second-class in conjunction with the gauge-fixing condition.  With this, we can clearly interpret the constraint $C$ as the overall Hamiltonian constraint that generates time evolution, and $\Delta$ as a gauge. 

Thus, the relation between the coupled systems and a system with a single time variable is obtained by interpreting one combination of constraints as gauge, and the other as the Hamiltonian constraint.  In turn, this is equivalent to a selection of the partial observables, which represent the quantities to which we assume we have access. 

The identification of $t_1$ and $t_2$ is not always necessary or convenient. For instance, in a general-covariant field theory like general relativity we have one Hamiltonian constraint per point of space, and a multi-fingered time evolution. In a sense, we can view the system as a coupling of one general-covariant system per each space point. A gauge fixing analogous to \eqref{gf} amounts then to reducing the multi-fingered time to a single global time. This can be convenient is some special situations where a global notion of time is easy to construct, but has no physical significance in general. 

Therefore in general the dynamics of the coupled system $S=S_1+ S_2$ is generally described by two \emph{independent} times. The choice of the partial observables in terms of which we want to describe the system is a matter of convenience, or additional physical inputs.   

A simple example of a physical input that can give rise to the gauge fixing just considered is an interaction $V_{\text{int}}$ that affects the systems at times $t_1$ and $t_2$ only when $t_1 =
t_2$.  

In what follows, we will see that a related additional physical input, which can select a particular combination of ``times" on physical grounds, is provided by specifying an equilibrium state of the coupled system. To do so, we need to move from mechanics to statistical mechanics.

\section{Statistical states and equilibrium}\label{see}

\subsection{Thermal time}

A statistical state $\rho$ is a (normalized)  smooth positive function on the phase space, which defines a statistical distribution in the sense of Gibbs \cite{gb}. In a conventional non-relativistic theory, where a Hamiltonian $H$ is given, the states of the form  $\rho \propto \exp{(-\beta H)}$ are the equilibrium thermal states, or Gibbs states. They are stationary and describe thermalized systems coupled to a thermal bath.   

Notice that (up to an overall multiplicative factor) the information on the time flow is coded into the Gibbs states as well as in the Hamiltonian. Indeed, a Gibbs state  $\rho$  is dual to the time flow $X_t$ in the sense that 
\be
\beta\rho \, \omega(X_t) = d\rho, 
\ee
where $\omega$ is the symplectic form.  This is a key equation. 

Thus, the time flow $\alpha_t$ can be recovered from the Gibbs state $\rho$ (up to the constant factor $\beta$,  discussed below in Sec. \ref{eqState}). This fact suggests that in a thermal context it may be possible to ascribe the dynamical properties of the system to the thermal state, rather than to the Hamiltonian: The Gibbs state determines a flow, and this flow is precisely the time flow. This idea, discussed in detail in \cite{ro93,roco} is the thermal-time hypothesis.\footnote{The phase-space symplectic structure  is the classical shadow of quantum non-commutativity. Therefore the state/flow relation can be traced to this non-commutativity \cite{roco}. Up to a unitary transformation, the flow is state-independent and therefore is only a property of the observable's algebra. This determines a remarkable intrinsic notion of time flow, where temporality is grounded in quantum non-commutativity.  Put naively: time is the difference between measuring $q$ and then $p$, and measuring $p$ and then $q$ \cite{cobook}.} 

Let us now come back to the general-covariant theory recalled in the previous section. Here one can define a {statistical state $\rho$} as a positive function $\rho:\Gamma_{ph}\to R^+$, normalized with respect to the Liouville measure $d\mu$, $(\int_{\Gamma_{ph}} d\mu \, \rho=1)$.
Any (non-degenerate) statistical state defines a vector field $X_{\rho}$ by
\be
\rho \,\sigma(X_{\rho})= d\rho, \label{fflow}
\ee
where $\sigma$ is the symplectic form on $\Gamma_{ph}$.  

The field $X_{\rho}$ generates a flow $\alpha^{\rho}_{\tau}$ on $\Gamma_{ph}$ called the \emph{thermal flow}; its generator 
\be
h= -\ln \rho \label{mod}
\ee 
is called the thermal Hamiltonian. The conjugate flow parameter $\tau$ is called \emph{thermal time} \cite{ro93,roco}.  For a non-relativistic system it is simply related to the non-relativistic time $t$ by 
\be
 \tau=\frac{t}\beta
\ee
where $\beta$ is the inverse temperature.  In mathematics, the thermal time flow is called the Tomita flow \cite{take}; the thermal hamiltonian is called the  \emph{modular Hamiltonian} in quantum field theory \cite{haag,lmr},  and the \emph{entanglement Hamiltonian} in the condensed matter context \cite{vidal}. For a recent discussion of this quantity in quantum gravity see also \cite{bimy}.

%
%

This thermal flow determines a time flow, i.e. a vector field $X_\tau$ on the constraint surface through the requirement 
\be
\label{flow}
\pi_{*} X_\tau =X_{\rho}.
\ee
 Solutions of the condition \eqref{flow} have a natural ambiguity due to the definition of the projection $\pi$: any of the vectors tangent to the orbits, the $Y_i$ of Sec. \ref{gcl}, project to zero and so if $X_\tau$ satisfies \eqref{flow} so does
 \be
 \tilde{X}_\tau = X_{\tau}+ \sum_{i=1}^D \alpha_i Y_i,
 \ee
 where the $ \alpha_i$ are arbitrary constants.  After illustrating the construction of $X_\tau$ we will show that this ambiguity plays no role in the determination of the dynamics. 

Consider again the harmonic oscillator described above: The statistical state $\rho=\exp{(-\beta H)} = \exp {(-\frac{1}{2}\beta A^2)}$ defines the thermal Hamiltonian $h=-\ln \rho = \frac{1}{2}\beta A^2$, whose vector field is $X_{\rho} = \beta \partial_{\phi}$. It follows immediately from \eqref{nn} that
\be
\pi_*(\partial_t)=-\partial_{\phi} \label{pii}
\ee
because
\be
\pi_*(\partial_t)=\frac{\partial A}{\partial t} \partial_A+\frac{\partial \phi}{\partial t} \partial_{\phi}=-\partial_{\phi}.
\ee
On the other hand, the kernel of the pre-symplectic form $\omega' = dp \wedge dq - dH \wedge dt$ is
\be
Y=\partial_t+\frac{\partial H}{\partial q} \partial_p-\frac{\partial H}{\partial p} \partial_q=\partial_t+q\,\partial_p-p\,\partial_q. \label{Y}
\ee 
So the general solution of (\ref{flow}) is\footnote{The unfortunate sign in the first term of this equation is due to the two convections that the oscillator dynamics is clockwise and that $\phi$ increases in the counter-clockwise direction.}
\be
X_\tau = -\beta \partial_t + \alpha Y.
\ee
 These relations are illustrated in the right panel of Fig.~\ref{uno}. The thermal flow (lower arrow) in the physical phase space determines a flow from orbit to orbit in the constraint surface.  

What to make of the ambiguity in the definition of $X_\tau$? As discussed in Sec. \ref{cl}, the orbits contained in the constraint surface code physical correlations between the partial observables of the system. However, these correlations are given in an unparametrized manner. In the case of a one-dimensional orbit all of the points of the curve $(t, q(t), p(t))$ are given as a set. The vector $Y$ is always tangent to this orbit but does not lead to a parametrization because $\omega'(Y)=0$. This shows that while the orbits capture all the physical correlations they do not encode dynamics. 

To understand the relation between this picture and the conventional description of time evolution, notice that the two solutions of the equation of motion
\be
q = A\,\sin(t + \phi),
\ee
and
\be
q = A\,\sin(t + (\phi+\delta\phi)),
\ee
differ \emph{only} in the fact that one is shifted in time with respect to the other.  This shows that a shift in time can be represented as a shift in the space of the solutions of the equations of motion. Dynamics is a motion from orbit to orbit. 

It is precisely this latter motion that is captured by the thermal flow $X_\rho$ and the physical symplectic form $\sigma$.  Remarkably, the time flow $X_\tau$ and the pre-symplectic form $\omega'$ capture  the same relations:
\be
\begin{aligned}
\omega'(\tilde{X}_\tau) &= \pi^* \sigma (\tilde{X}_\tau) = \sigma(\pi_* \tilde{X}_\tau) = \sigma (\pi_* X_\tau) \\
 & = \sigma( X_\rho). 
\end{aligned}
\ee
Thus the ambiguity in the determination of $X_\tau$ has no impact on the dynamics generated by this vector field. 

This completes the covariant treatment. In special cases it is also possible to identify the partial observable $t$ acting as time. This is achieved by making a split of all the partial observables into one, call it $t$, and the rest and then finding an $X_\tau$ such that 
\be
i_* X_\tau = X_t|_{\cal C},
\ee
where $X_t \propto \partial_t$ and  $X_t|_{\cal C}$ denotes the restriction of this vector field to the constraint surface; as always, $\partial_t$ means the vector field where $t$ is changed while the other partial observables coordinatizing $\Gamma_{ex}$ are held fixed. This procedure relies on a nice choice of partial observables from the outset and may be difficult to carry out in general, however it clarifies the recovery of the usual formulation in simple examples such as the oscillator discussed here. The relationships are summarized by
\be
X_t|_{\cal C}  \xleftarrow{\,\,i_*\,} X_\tau \xrightarrow{\,\pi_*\,} X_{\rho}. \label{time}
\ee

In conclusion, the general-covariant description of the dynamics of a system treats the time variables exactly on the same ground as the other variables.  But the selection of a state distinguishes a flow, the thermal flow. In particular, a Gibbs state $\sim e^{-\beta H}$ singles out the variable $t$ (possibly scaled by $\beta$) as the thermal time. We can now use these tools in the context of coupled covariant systems.

\subsection{Equilibrium state and common time}
\label{eqState}

Consider two systems $S_1$ and $S_2$ with (extended) phase space $\Gamma_{ex}=\Gamma_{ex}^1\times \Gamma_{ex}^2$, and constraints $C_1$, $C_2$. {A  statistical state of the combined system} is defined by a probability distribution $\rho$ on the physical phase space of the coupled system 
\be
\rho : \Gamma_{ph}^1 \times \Gamma_{ph}^2\to R^+ \label{pro}.
\ee
An  \emph{equilibrium state} should be factorizable into the product of two states \cite{ll}. Let us thus consider a state of the form 
\be
\rho=\rho_1 \cdot \rho_2. \label{facto}
\ee
where $\rho_i$ is a function on $\Gamma_{ex}^i$.

Now, let $X_{\rho}$, be the thermal flow generated by this state. Its generator can be decomposed into the sum of two components $X_{\rho}^1$ and $X_{\rho}^2$, with components each in one of the two factor spaces. From the factorization property (\ref{facto}) it follows that the Lie brackets of $X_{\rho}^1$ and $X_{\rho}^2$ vanish 
\be
[X_{\rho}^1,X_{\rho}^2] =0.\label{lie}
\ee 
Equivalently,
\be
X_{\rho}^2(\rho_1)=0 \quad \text{and} \quad X_{\rho}^1(\rho_2)=0.
\ee
The thermal Hamiltonian of the coupled system is a sum of two terms, each one defining a flow in its own physical phase space, but notice that the statistical state $\rho$ defines a \emph{common} flow, with a single time for the two systems. This fixes a  preferred split of the system of constraints into a single global Hamiltonian constraint and a gauge. 

Again consider the case of two harmonic oscillators: Let the equilibrium statistical state of the coupled system be
\be
\rho = e^{-\frac{1}{2}(\beta_1{A_1}^2+\beta_2{A_2}^2)}. \label{product}
\ee
The Hamiltonian vector field $X_{\rho}$ is defined by
\be
\sigma(X_{\rho})= d \ln \rho. \label{floww}
\ee
Now, in the coupled system, the symplectic form on $\Gamma_{ph}$ is 
\ba
\sigma&=&\sigma_1 + \sigma_2 \\ \nonumber
&=& A_1\,\text{d}A_1 \wedge \text{d}\phi_1+A_2\,\text{d}A_2 \wedge \text{d}\phi_2.
\ea
Equation (\ref{floww}) then reads
\ba
\sigma_1(X_{\rho}^1) + \sigma_2 (X_{\rho}^2)&=& \text{d} \ln \rho_1 +\text{d} \ln \rho_2 \\ \nonumber
&=&-\beta_1\,A_1\,\text{d}A_1- \beta_2\,A_2\,\text{d}A_2,\label{ortoflow}
\ea
and is solved by\footnote{Helpful for dimensional analysis: the symplectic form is $\sigma=m\omega^2A\,\text{d}A \wedge \frac{\text{d}\phi}{\omega}$. The statistical state is $\rho=e^{-\frac{1}{2}\beta (m\omega^2 A^2)}$; therefore  $\sigma(X_{\rho})=-\beta\, m\, \omega^2A\, \text{d}A$. Consistently, we have $X_{\rho}=\beta (\omega \partial_{\phi})$. Indeed, multiplying by $\hbar$ we would get a dimensional flow, as expected for the thermal time flow.}
\be
X_{\rho}=\beta_1\partial_{\phi_1} + \beta_2\partial_{\phi_2}. \label{coflow}
\ee
The time vector fields of the two subsystems, $X_{\rho}^1\equiv \beta_1\partial_{\phi_1}$ and $X_{\rho}^2 \equiv \beta_2\partial_{\phi_2}$, satisfy the equilibrium condition in (\ref{lie}).

Now, following the argument in (\ref{time}), the time flow on the constraint surface of the combined extended phase space is generated by the vector field
\be
X_\tau=\beta_1\partial_{t_1} + \beta_2\partial_{t_2}, \label{cocoflow}
\ee
as a sum of two \emph{independent} components.

Therefore, the time function on the combined extended phase space is expressed in terms of the coupled system thermal time $\tau$, 
\be
t=\beta \,\tau = \beta\,\left(\frac{t_1}{\beta_1}+ \frac{t_2}{\beta_2}\right). \label{truetime}
\ee 
The variable $t$ parametrizes the orbit on which the thermal Hamiltonian $h=\frac{1}{2}({\beta_1A_1}^2+{\beta_2A_2}^2)$ is conserved, while the factor $\beta$ indicates how fast this orbit is traversed. 

At first, the idea of different temperatures and a mixed time may sound implausible, but a simple example will convince the reader that the physics is right. 

Imagine that the two harmonic oscillators of the example are placed at different levels in a static gravitational field and exchange heat with a common thermal bath (a warm gas, or radiation field, in which they are immersed).  Because of the Tolman effect \cite{tolman}, namely because of the gravitational redshift, they will not be at the same local temperature: the one lower will be at higher temperature.  Therefore $\beta_1\ne\beta_2$. The thermal time variable  \eqref{truetime} is neither of the two local proper times along the oscillators' worldlines, which govern their local dynamics: rather, it is the global Killing time with respect to which the metric is static and therefore equilibrium can be established.  It is immediate to see that the time defined in \eqref{truetime} is precisely this Killing time (which is also defined up to a global rescaling, like the thermal time). The ratio between the two local times in  \eqref{truetime}, implied by the different temperatures $\beta_1$ and  $\beta_2$ is nothing else than the Tolman law. 

Summarizing, the statistical state in (\ref{product}) singles out a preferred time variable for the coupled system. The two time vectors on the constraint surfaces are now related. 

By choosing a time variable $t$, that is a specific combination of $t_1$ and $t_2$, the statistical state  defines the particular combination of $C_1$ and $C_2$, that we then interpret as the overall Hamiltonian constraint for the coupled dynamics.

A dual perspective where common time is split into two, local times is discussed in \cite{Menicucci}.

\section{Equilibrium and vanishing information flux}

Once a common time variable $t$ is introduced via (\ref{product}), we can re-coordinatize $\Gamma_{ex}=\Gamma_{ex}^1\times \Gamma_{ex}^2$ inorder to have a symplectic form $\omega$ defined in terms of ``new'' canonical common variables $(t, \chi, E_t, E_{\chi})$, such that \\
\ba
&&\{t,E_t\}=1, \ \  \{t,\chi\}=0; \ \ \label{pit} \{t,E_{\chi}\}=0;\\ 
&&\{\chi,E_{\chi}\}=1;\ 
\{\chi,E_t\}=0.\label{gauged}
\ea

In particular, starting from (\ref{truetime}), one can use (\ref{pit}) to define $E_t\propto \frac{1}{2\beta}(\beta_1\,E_1+\beta_2\,E_2)$, and (\ref{gauged}) to define $\chi\propto \beta\,\left(\frac{t_1}{\beta_1}- \frac{t_2}{\beta_2}\right)$. 

With an explicit expression for $\chi$, we can reinterpret the gauge fixing condition (\ref{syncro}) within the statistical approach: setting $\chi=const$ implies
\be
\frac{ t_1}{\beta_1}-\frac{ t_2}{\beta_2}=const,
\ee
which can be rewritten in terms of thermal time intervals, 
\be
d \tau_2 =d\tau_1. \label{thermeq}
\ee
This is the relation between the local thermal times defined by an equilibrium state. How to interpret this result?

\begin{figure}[t]
\includegraphics[width=0.30 \textwidth]{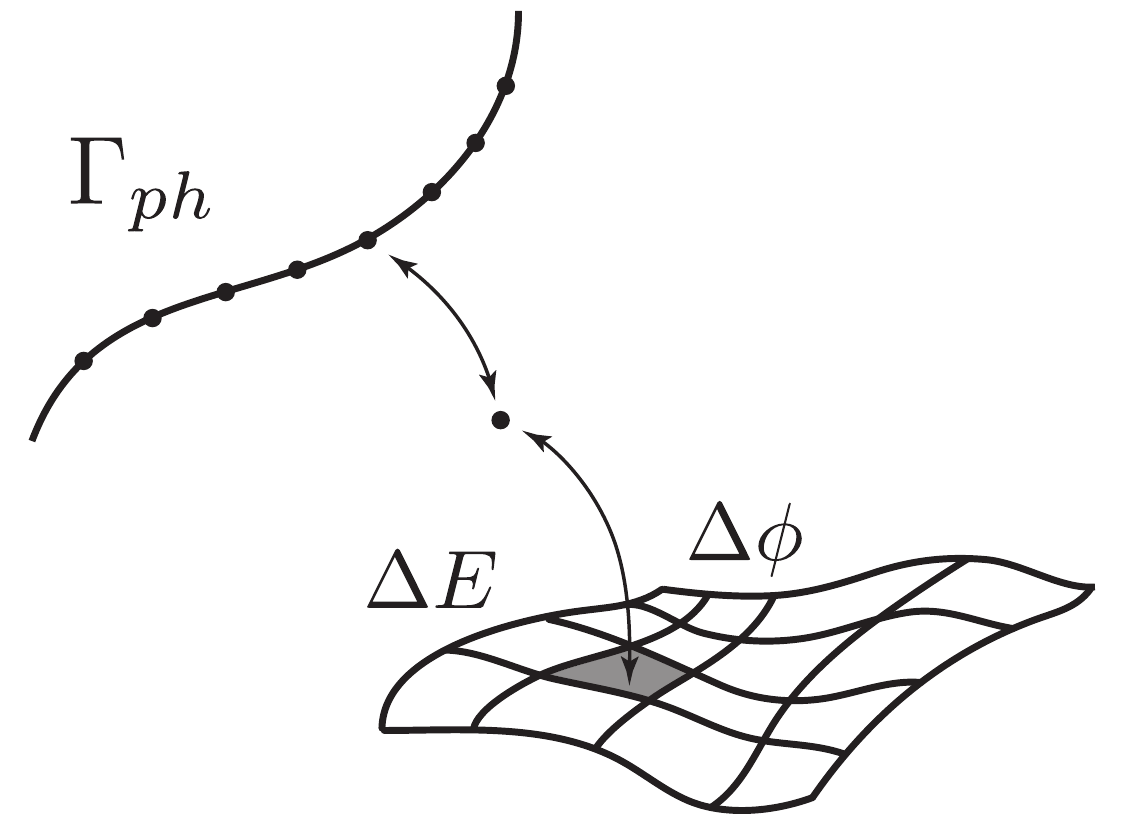}
\caption{The steps along an \emph{history} can be seen as a sequence of phase space regions, each with area $\sim h$.}\label{history}
\end{figure}

The answer is in \cite{hal}, whose argument we include here for completeness.
The thermal time $\tau$ parametrizes a curve defined by a sequence of physical states on $\Gamma_{ph}$. Along this \emph{history} of states, one can think of thermal time $\tau=t/\beta$ as time measured in terms of elementary ``time steps'', where a step is the characteristic time taken to move to a distinguishable cell in the phase space at a given temperature (see Fig. \ref{history}). 

Now, consider two systems that are coupled via some interaction during a certain interval. During the interaction interval the first system transits $N_1$ states, and the second $N_2$. Since an interaction channel is open, each system has access to the information\footnote{Information is meant here as a measure of a number of states, as it is defined in the classic text by Shannon \cite{sha}.} about the states the other has transited via the physical exchanges of the interaction.

If system 2 has access to an amount of information $I_1 = \log N_1$ about system 1, and system 1 has access to an amount of information $I_2 = \log N_2$ about system 2, then the net flow of information can be defined as $\delta I=I_2-I_1$. For the equilibrium regime we are considering, one can postulate that, as with any other flow, also information flow $\delta I$ must vanish, namely
\be 
\delta I = 0, \label{infoeq}
\ee
or, equivalently, 
\be
N_1 = N_2.
\ee
In particular, since the rate that states are transited is given by $\tau$ and we assume a fixed interaction interval, the equilibrium conditions also reads
\be
\tau_1= \tau_2.
\ee

\begin{figure}[t]
\includegraphics[width=0.3 \textwidth]{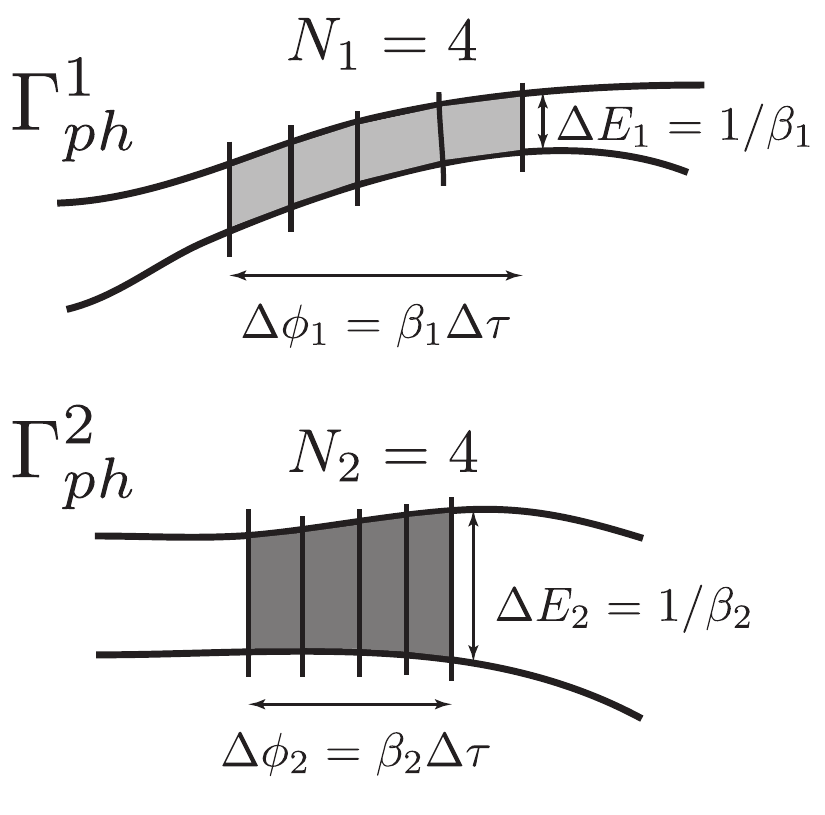}
\caption{In \cite{hal} it was shown that equilibrium between histories occurs when they transit the same number of Planck cells during the same interval, that is when $N_1=N_2$. This figure demonstrates that the same condition holds when the thermal time derived from a statistical state is used.}\label{pps}
\end{figure}

Once again we illustrate this with a harmonic oscillator, consider the statistical state $\rho \sim e^{-\frac{1}{2}\beta A^2}$: the Hamiltonian $H=\frac{1}{2} A^2$ has both mean value $E=\langle H\rangle =\frac{1}{\beta}$ and variance $\Delta E =\frac{1}{\beta}$, while the variable $\phi$ is spread. Suppose we consider a lapse $\Delta \tau$ of common time.
Then $\Delta t_1 = \beta_1\, \Delta \tau$ and $\Delta t_2 = \beta_2 \, \Delta \tau$. This can be viewed as a motion along the orbits on the constraint surface , or, equivalently, as a change of state $\Delta \phi_1 = \beta_1 \, \Delta \tau$ and $\Delta \phi_2 = \beta_2\, \Delta \tau$ (see Fig. \ref{pps}). If we fix $\Delta \tau$, the number of Planck-sized cells covered by the two systems is, respectively,
\ba
N_1=\Delta \phi_1\Delta E_1=\beta_1\, \Delta \tau / \beta_1=\Delta \tau\\
N_2=\Delta \phi_2\Delta E_2=\beta_2\, \Delta \tau /\beta_2=\Delta \tau.
\ea
Therefore, we get
\be
N_1 =N_2. \label{Ninfo}
\ee

This is precisely the general covariant condition for equilibrium in terms of zero-information flux introduced in \cite{hal}.  The gauge fixing in (\ref{thermeq}) is equivalent to the equilibrium condition in (\ref{Ninfo}). In this sense, it is implicitly fixed in the statistical approach, where equilibrium is assumed. 

Summarizing, a state which is factorized defines a preferred flow of time with respect to which the two systems are in equilibrium. The common flow of time does not need to agree with an independent local notion of proper time defined for each system individually, as the example of the temperature dependence on the gravitational potential shows.  Equilibrium singles out one combination of local times as the thermal time in which thermalization occurs. In doing so, it breaks the equivalence between all the constraints, singling out a combination of these as the overall Hamiltonian constraint.

\section{Conclusions}\label{disc}


We have shown that the coupling of general-covariant systems gives rise naturally to systems where the dynamics is coded into a system of constraints. This system lacks a preferred notion of time.  A preferred notion of time is selected by a statistical state.  For coupled systems, a factorized statistical state breaks the equivalence between the constraints and singles out a common time flow.  With respect to this time flow, the systems are in equilibrium, and equilibrium is characterized by the vanishing of the information flux, as suggested in \cite{hal}.

Here we have started from a \emph{given} statistical state.  Of course, one can ask what are the physical conditions for the realization of one or another of these states.  These can probably be related to the thermalizing interaction (which we have not yet considered) between the two systems.  For instance, two non-relativistic systems exchange energy via heat at times $t_1=t_2$; while if $t_1$ and $t_2$ are the proper times of the systems this is not anymore true for two systems at different levels in a gravitational potential. The relation between the thermalizing interaction and the state will be studied elsewhere. 

An intriguing relation between equilibrium and gauge emerges from this analysis.  The constraints that are not identified as the overall Hamiltonian constraint play the role of gauge generators.  The corresponding gauge refers to partial observables that are irrelevant for the thermal description of the coupled system.  This picture appears to  support the interpretation of gauge invariance discussed in \cite{gauge}: gauge as a description of partial observables that cannot be predicted but can be measured in interactions with external systems, rather than just mathematical redundancy.  The ``thermal gauge" degrees of freedom do not play a role in the statistical description of the coupled system. They are somehow like microscopic variables neglected in a coarse graining.   A Boltzmann thermalizing interaction conserves a minimal number of physical quantities, but averages over the information on any other observables. The choice of a statistical state for the coupled system singles out an overall Hamiltonian constraint, and reduces irrelevant partial observables to gauges.

\subsection*{Acknowledgments}
We are grateful to Aldo Riello for discussions related to this paper. HMH acknowledges support from the National Science Foundation (NSF) International Research Fellowship Program (IRFP) under Grant No. OISE-1159218.

\bibliographystyle{apsrev4-1}

\end{document}